\documentclass[journal=apchd5,manuscript=article, layout=twocolumn]{achemso}

\usepackage[version=3]{mhchem} 
\usepackage{siunitx}

\author{Huili Hou}
\affiliation{Quantum Engineering Centre for Doctoral Training, H. H. Wills Physics Laboratory and Department of Electrical and Electronic Engineering, University of Bristol, Tyndall Avenue, BS8 1FD, UK}
\alsoaffiliation{Quantum Engineering Technology Labs, H. H. Wills Physics Laboratory and Department of Electrical \& Electronic Engineering, University of Bristol, BS8 1FD, UK}
\alsoaffiliation{currently with NuQuantum, Broers Building, JJ Thomson Ave
Cambridge, CB3 0FA}
\author{David Dlaka}
\affiliation{Quantum Engineering Technology Labs, H. H. Wills Physics Laboratory and Department of Electrical \& Electronic Engineering, University of Bristol, BS8 1FD, UK}
\email{david.dlaka@bristol.ac.uk}
\author{Jon Pugh}
\affiliation{Quantum Engineering Technology Labs, H. H. Wills Physics Laboratory and Department of Electrical \& Electronic Engineering, University of Bristol, BS8 1FD, UK}
\author{Ruth Oulton}
\affiliation{Quantum Engineering Technology Labs, H. H. Wills Physics Laboratory and Department of Electrical \& Electronic Engineering, University of Bristol, BS8 1FD, UK}
\author{Edmund Harbord}
\affiliation{Quantum Engineering Technology Labs, H. H. Wills Physics Laboratory and Department of Electrical \& Electronic Engineering, University of Bristol, BS8 1FD, UK}
\title{Nanoring Tamm Cavity in the Telecommunications O band}


\begin{document}

\begin{tocentry}

\begin{center}
\includegraphics[width=1\textwidth]{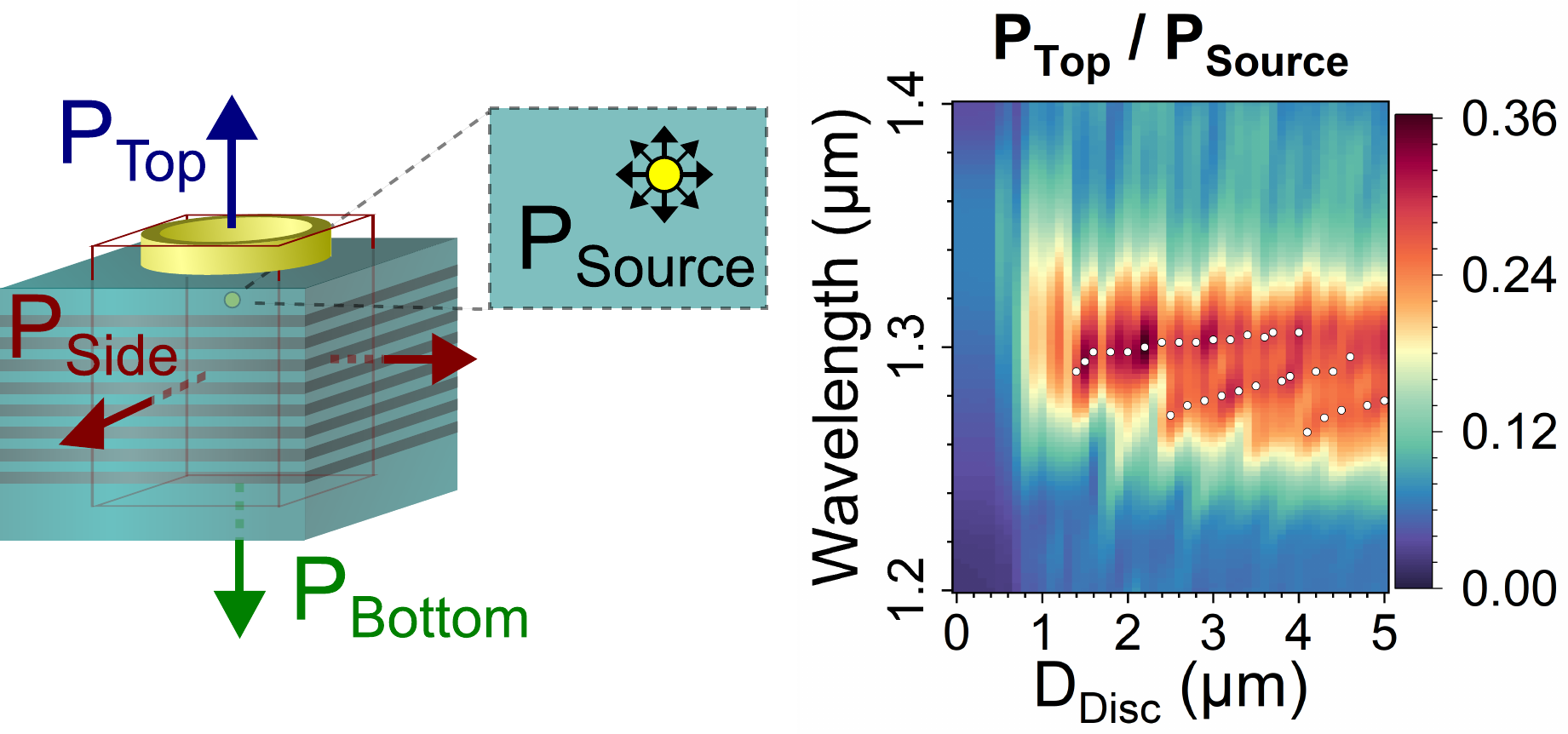}
\end{center}

\end{tocentry}

\begin{abstract}
Quantum and classical telecommunications require efficient sources of light. Semiconductor sources, owing to the high refractive index of the medium, often exploit photonic cavities to enhance the external emission of photons into a well-defined optical mode. Optical Tamm States (OTS), in which light is confined between a distributed Bragg reflector and a thin metal layer have attracted interest as confined Tamm structures are readily manufactureable broadband cavities. Their effficiency is limited however by the absorption inherent in the metal layer. We propose a nanoring Tamm structure in which a nanoscale patterned annular metasurface is exploited to reduce this absorption and thereby enhance emission efficiency. To this end, we present designs for a nanoring Tamm structure optimised for the telecommunications O band, and demonstrate a near doubling of output efficiency ($35\%$) over an analogous solid disc confined Tamm structure ($18\%$). Simulations of designs optimised for different wavelengths are suggestive of annular coupling between the Tamm state and surface plasmons. These designs are applicable to the design of single photon sources, nanoLEDs, and nanolasers for communications.    

\end{abstract}

\section*{Introduction}
A high-quality single photon source (SPS) with high efficiency, indistinguishability, and purity is an essential component for quantum information applications such as high-rate boson sampling \cite{wang2017b}, quantum computing \cite{varnava2008,obrien2007} and quantum communication \cite{vajner2022}. A quantum dot (QD) - a nanoscale inclusion of one semiconductor embedded within another - is an excellent candidate for such a single photon source, as it exhibits an internal quantum efficiency close to 100\%. However, as the QD is embedded within a semiconductor host matrix with a high refractive index, this poses a challenge in extracting the photon and limits the external quantum efficiency of the SPS.
A number of photonic structures to enhance the external efficiency have been proposed and demonstrated such as pillar microcavities\cite{wang2017b, somaschi2016, wang2019b}, photonic crystal cavities \cite{madsen2014}, waveguides \cite{uppu2020a}, gratings \cite{wang2019b, liu2019}, and solid immersion lenses \cite{chen2018}. They enhance the interaction between the light field and the emitter, and efficiently funnel the photons into an optical mode. However, all of these structures require exquisite fabrication at the nanoscale, and these demanding fabrication techniques often introduce defects into the semiconductor close to the QD, which can impact the quality and brightness of the emitted photons.

An alternative is the Tamm structure, a photonic cavity in which light is confined within a Distributed Bragg Reflector (DBR) terminated by a thin metal layer \cite{Kaliteevski2007}. The photonic stop-band of the DBR and the negative permittivity of the metal confine light in the vertical direction. The planar Tamm structures can be modified to obtain an additional lateral confinement by patterning the metal layer into a disc with a finite diameter \cite{gazzano2011, Parker}  forming a confined Tamm structure which has attracted considerable interest for SPS applications \cite{gazzano2011, Parker, gazzano2012, parker2018}. In such structures, the light is confined in Optical Tamm States (OTS), which are sometimes referred to in the literature as "Tamm plasmons", by analogy with the surface plasmon-polaritons (SPP) that can form at a metal-semiconductor interface. Unlike true SPPs however, the electric field maximum in a Tamm state lies within the dielectric rather than at the metal-dielectric interface, reducing ohmic losses in the metal. Furthermore, light couples into these optical Tamm states without the need for a prism or a grating to achieve momentum matching, unlike SPPs.

Compared with the more widely-used pillar microcavities, Tamm-based devices benefit from a simple fabrication process that requires only the deposition of a single layer of metal and does not require etching of the semiconductor, thereby avoiding the introduction of defects that will degrade the photon emission.
Moreover, the low quality factor inherent in a confined Tamm state has the potential for higher device yield, owing to the less stringent requirement for spectral overlap of the emitter and spectrally broad cavity \cite{gazzano2012, lundt2016, harbord2019}.  However, the quantum efficiency from the structure is limited due to the scattering loss and absorption in the metal \cite{gazzano2012} and we look for routes to decrease this absorption mechanism. 

For quantum telecommunication applications \cite{arakawa2020}, it is essential to design an SPS working at the telecom O band or C-band, where the transmission loss in optical fibres is minimised. In previous work, an OTS centered in the telecom O band has been designed \cite{parker2018} and fabricated \cite{Parker,harbord2019}.  In this paper, we further optimize such confined Tamm structures for telecom O- band (around 1.3$\mu$m) and investigate the effect of introducing an aperture into a confined Tamm structure. 

Aperture-in-metal structures have been actively studied since 1944 \cite{Ebbesen1998, lezec2002, degiron2004, garcia-vidal2010, pugh2014} and used to improve the performance of photonic devices \cite{Mehfuz2004, gordon2018, shafiq2022, wang2022}. For example, metallic nanoring structures can be used to extract photons from a high refractive index dielectric with high efficiency \cite{trojak2017, haws2022}. The mechanisms for this are complex, and arise due to the coupling of light to the localized optical modes formed by the aperture, such as diffraction modes and plasmonic resonance modes. Coupling between a Tamm mode and these localized modes has been demonstrated recently \cite{Afinogenov2013, lopez-garcia2014, Azzini2016, symonds2017}. By etching a periodic grating into a Tamm structure, the SPP-Tamm coupling controls the far-field of the emission \cite{lopez-garcia2014}. Exploiting metal ring/dielectric “super Tamm” structures \cite{symonds2017} has allowed dramatically higher Q factors than are usually possible with confined optical Tamm states. 

 By using Finite-Difference Time-Domain (FDTD) simulations, we show that these nanoring Tamm cavities can reduce the absorption in the metal and as a result improve the extraction efficiency by a factor of almost two, while maintaining the broadband emission possible with low Q factors and maintaining the ease of fabrication associated with Tamm structures. 

This improvement is attributed to the coupling between Tamm mode and the local aperture mode.  This strategy can potentially be applied to enhance the emission efficiency of a variety of Tamm-based devices including light emitting diodes\cite{Zheng2019, pugh2021, SaruaHarbord2023},  narrow-bandwidth thermal emitters \cite{Yang2017}, lasers\cite{Symonds2013, Toanen2020} and single photon sources \cite{gazzano2011, Parker, gazzano2012, parker2018}. 

\section*{Methodology}

\begin{figure*}
\includegraphics[width=0.65\linewidth]{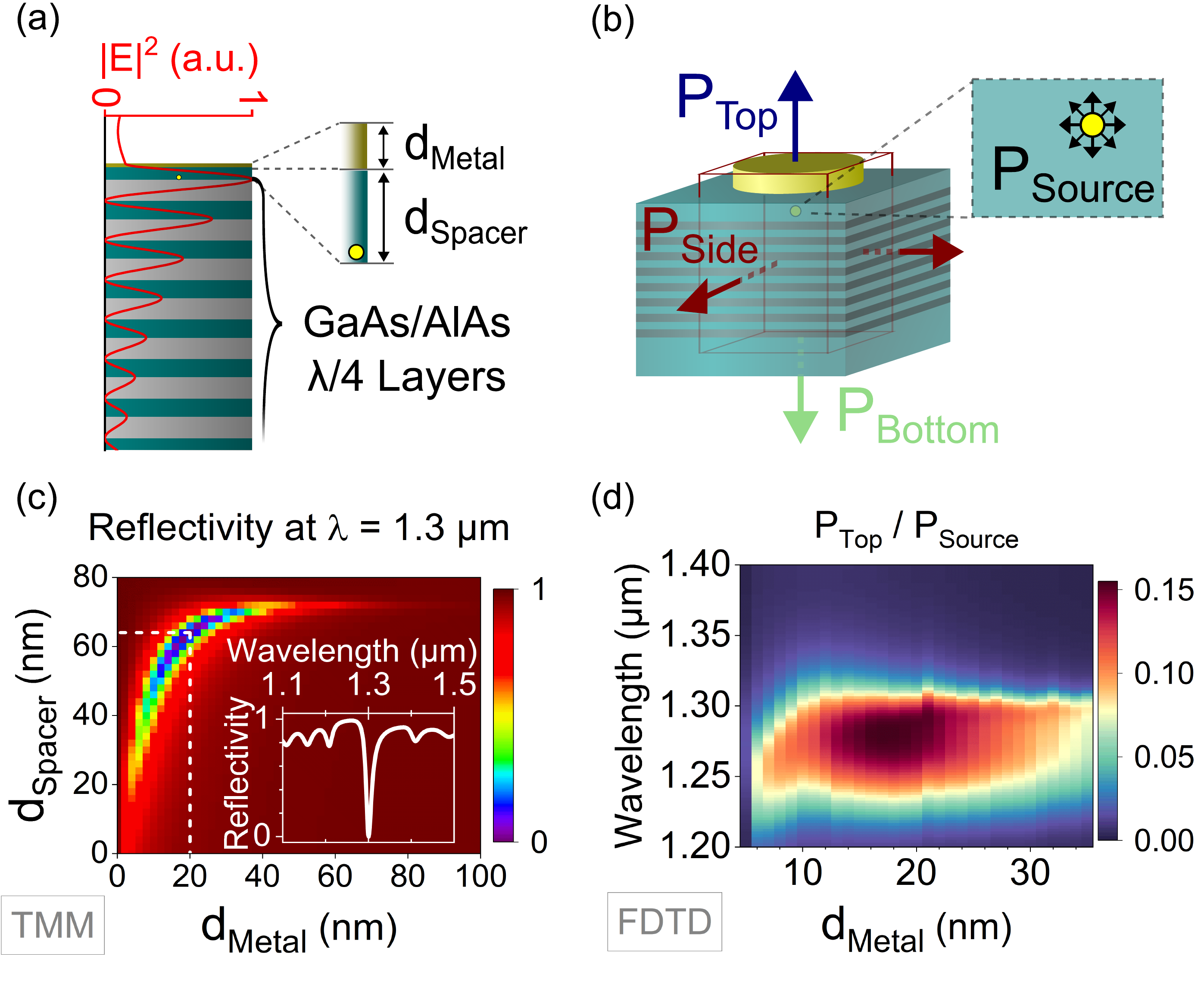}
\caption
{
(a) A 1D schematic illustration of a Tamm structure, consisting of a thin layer of metal, a high refractive index spacer layer, and a distributed Bragg reflector (DBR). The E field corresponding to the Optical Tamm State, as calculated by the Transfer Matrix Method, is shown in red. To ensure maximal source-cavity coupling, the emitter is positioned a few nanometers above the interface, close to the E-field maximum. (b) A schematic diagram illustrating the Finite Difference Time Domain (FDTD) model, with a source emitting $P_\text{Source}$ power, and a cuboid bounding the structure so that power flux through the 2D surfaces can be calculated. (c) TMM calculations of the reflectivity at $\lambda = 1.3\mu$m  for a range of spacer and metal layer thicknesses, with the inset showing the reflectivity as a function of wavelength corresponding to $d_\text{Spacer} = 64$ nm and $d_\text{Metal} = 20$nm. The dip in reflectivity signifies coupling to a cavity, and reveals the set of specific spacer and metal layer dimensions which result in the desired Optical Tamm State wavelength. (d) The FDTD simulated efficiency of source power which couples out of the top of the structure as a function of $d_\text{Metal}$ (where $d_\text{Spacer}$ is appropriately modified to keep the Optical Tamm State centred at $1.3\mu$m). The out-coupling efficiency peaks at 15-16\% for 20nm of gold.}
\label{fig1}
\end{figure*}

An Optical Tamm State forms when an emitter resonates inside a high refractive index layer, referred to as the spacer layer with a thickness $d_\text{Spacer}$, which is bound by a DBR on the bottom and a thin metal film of thickness $d_\text{Metal}$ on top. A Transfer Matrix Method (TMM) calculation of the 1.3$\mu$m E-field has been presented in Fig.\ref{fig1}(a) in red, superimposed on an illustration of a 1D Tamm structure. The quantum dot is positioned in the spacer layer, a few nanometers above the interface with the $\lambda/(4n_\text{Spacer})$ DBR where the E-field peaks. Building on previous work\cite{Parker, parker2018}, we use 17.5 DBR pairsterminated by a gold metal layer. We use TMM to initially probe the structural parameters, while FDTD simulations (using a commercial solver) provide a more comprehensive 3D treatment which appropriate for confined (radially finite) structures. We calculate the power transmitted through the top, side, and bottom faces of a cuboid which circumscribes the Tamm structures ($P_\text{Top}$,$P_\text{Side},$ and $P_\text{Bottom}$, respectively) as illustrated in Fig.\ref{fig1}(b), and normalise to the power emitted by the dipole source, $P_\text{Source}$ such that the absorbed power $P_\text{Absorbed}$ in the metal layer is given by

\begin{equation}
        P_\text{Absorbed} = P_\text{Source} - (P_\text{Top} + P_\text{Side} + P_\text{Bottom})
    \label{AbsEq}
\end{equation}
with the absorption $\alpha$ being
\begin{equation}
        \alpha = \frac{P_\text{Absorbed}}{P_\text{Source}}
\end{equation}
To further contextualise the device performance, we calculate the cavity-emitter active coupling $\beta$ as

\begin{align}
\beta &= \Gamma_\text{OTS}/(\Gamma_\text{OTS}+\gamma)\\
  &=\frac{P_\text{Top}+P_\text{Bottom}}{P_\text{Top}+P_\text{Side}+P_\text{Bottom}}
\end{align}

where the decay rate into the vertical cavity mode i.e. the Optical Tamm State is given by $\Gamma_\text{OTS}$, and $\gamma$ is the spontaneous decay rate in bulk. The passive efficiency $\eta$ and the efficiency at the first lens $\xi$ are given by

\begin{align}
\eta &=\frac{P_\text{Top}}{P_\text{Top}+P_\text{Bottom}} \\
\xi &= \beta \eta (1-\alpha) = \frac{P_\text{Top}}{P_\text{Source}}
\end{align}

The $\beta$ factor accounts for losses at the sides, the $\eta$ factor for leakage into the substrate, and $\alpha$ for the absorption in the metal film, such that $\xi$ corresponds to the fraction of photons that can be extracted from the Tamm structure. The cavity resonance forms at the frequency where the metal and spacer impedances are matched and consequently, changes in either $d_\text{Spacer}$ or $d_\text{Metal}$ result in changes to the OTS energy \cite{Vinogradov2006,Kaliteevski2007}.

A TMM calculated 2D plot of the reflectivity at $\lambda=1.3\mu$m for a wide range of both spacer ($d_\text{Spacer}$) and metal layer ($d_\text{Metal}$) thicknesses is shown in Fig.\ref{fig1}(c). A reduction in the reflectivity corresponds to coupling to an Optical Tamm State, as illustrated in the figure inset. For each $d_\text{Metal} \approx 10-40$nm value, there is a corresponding $d_\text{Spacer}$ in the $20-70$nm range which results in an OTS at 1300nm, and these two parameters are no longer independent. This reduces the dimensionality of the structural optimisation of the $d_\text{Spacer}$ and $d_\text{Metal}$, and removes any unwanted effects due to spectral detuning between the OTS and the $1.3\mu m/(4n)$ DBR. The internal efficiency, corresponding to $P_\text{Top}/P_\text{Source}$, obtained from 3D FDTD simulations of a 1D Tamm (where the metal layer is semi-infinite and has no radial boundary) has been shown in Fig.\ref{fig1}(d) as a function of wavelength and metal thickness, while $d_\text{Spacer}$ is set to keep the OTS wavelength constant at 1.3$\mu$m. As the Tamm states are not laterally confined, the emission is highly multimoded which contributes to a broadband emission efficiency. This is not the case when a lateral confinement is introduced in the following section. It can be seen that if $d_\text{Metal} < 10$nm, the gold film is too thin to impedance match the 1250-1350DBR. If on the other hand $d_\text{Metal} > 30$nm, losses due to absorption and increased reflectivity become too large. The efficiency of $\lambda = 1.3\mu$m is maximised at $d_\text{Metal} = 20$nm, with a corresponding $d_\text{Spacer} = 64$nm; we choose these values for the layer thicknesses moving forward. 

Simulations related to the planar structure of maximum efficiency are shown in figure 2. Without the metal, most light ($\sim$90\%) travels along the DBR, while little ($\sim$5\%) light is emitted vertically. On introduction of 20nm gold, the vertically emitted light at the design wavelength is increased ($\sim$10\%) with a concomitant increase in the absorption ($\sim$ 30\%). In the rest of this work we show how to reduce the absorption.

\begin{figure}
\includegraphics[width=\linewidth]{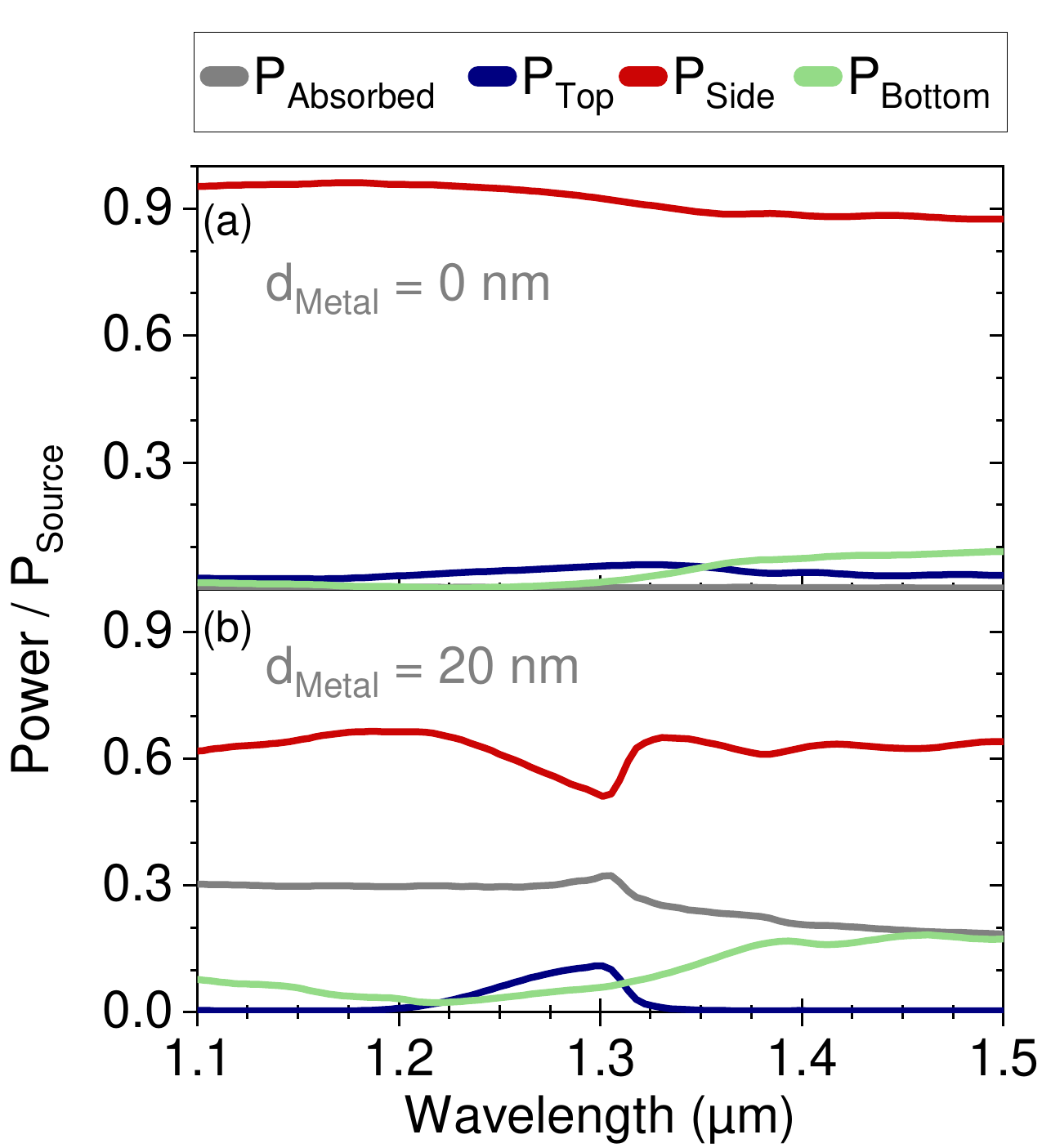}
\caption
{
FDTD simulations of a planar Tamm structure as illustrated in Fig.\ref{fig1}(a) showing power transmission through the relevant cuboid surfaces. The absorbed power in the metal $P_\text{Absorption}$ (grey) and the power transmission through the top (blue), side (red), and bottom (green) of the Tamm structure have been normalised to the total power produced by the dipole emitter $P_\text{Source}$. Calculations are shown for a 17.5 pair DBR with a 64 nm spacer layer with (a) no metal on the top (b) a metal layer 20nm thick to form a Tamm state at 1.3$\mu$m as described in the text. We note that both the emission though the top $P_\text{Top}$/$P_\text{Source}$ and the absorption are enhanced by the formation of the Optical Tamm State.
}
\label{fig2}
\end{figure}

\section*{Minimising Losses for Confined Optical Tamm States}

\begin{figure*}
\includegraphics[width=\linewidth]{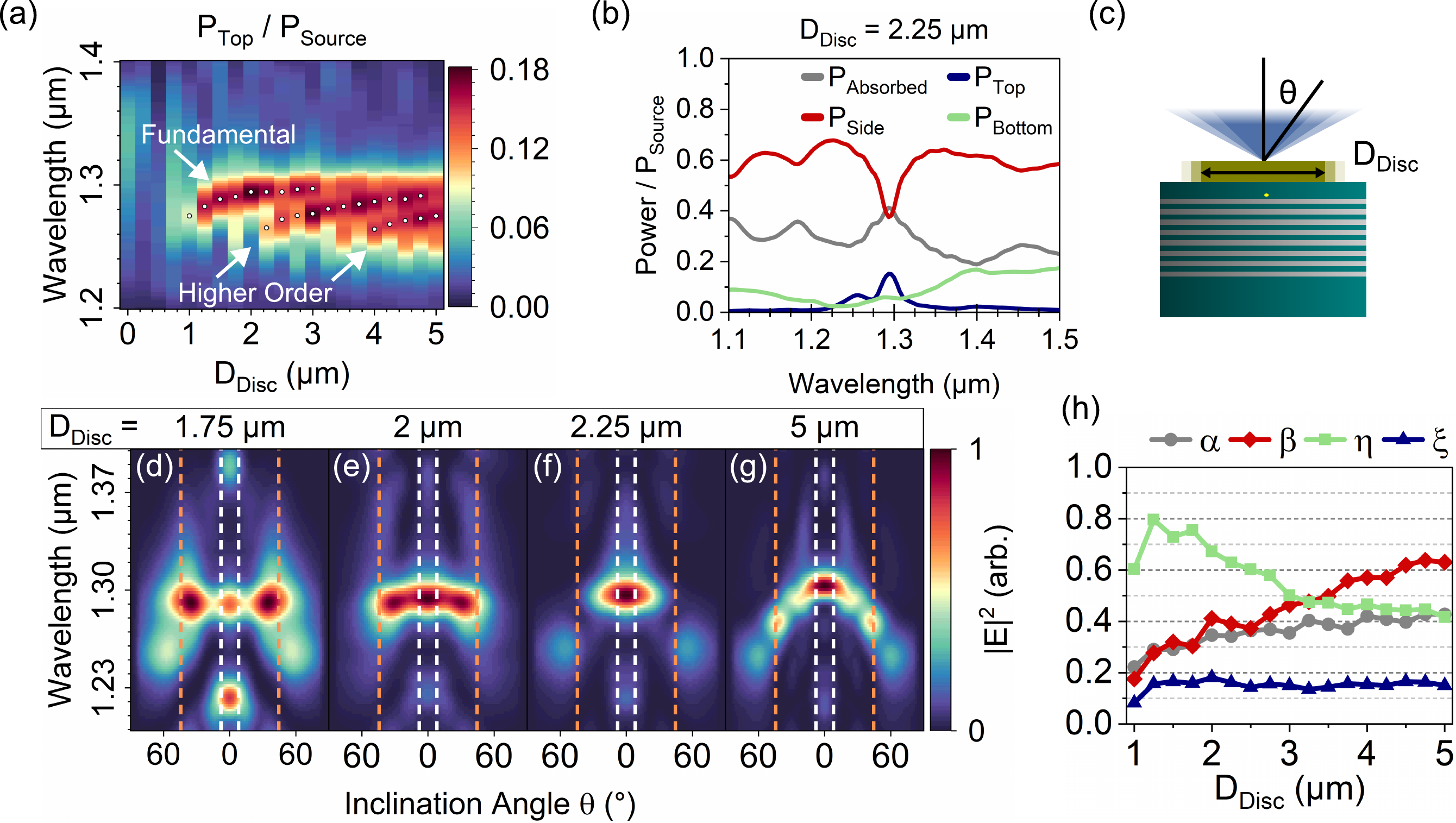}
\caption
{
Summary of FDTD simulations of a confined Tamm state due to a disc of finite diameter $D_\text{Disc}$. (a) The power transmission though the top of the structure ($P_\text{Top}/P_\text{Source}$) as a function of disc diameter. Local maxima are indicated with white points, serving as a guide to the eye for both the fundamental and higher order modes. (b) Absorption and power transmission as a function of wavelength for a disc of diameter $D_\text{disc}= 2.25 \mu m $. (c) Schematic illustrating the disc diameter variations and the corresponding effect of the far field distribution as a function of inclination angle $\theta$. (d)-(g) The far field distribution of E-field as a function of inclination angle $\theta$ calculated for a range of disc diameters. Numerical apertures of an SMF-28 fibre (NA = 0.14, white dotted lines) and a typical microscope objective (NA = 0.70, orange dotted lines) are marked in each case. (h) The absorption $\alpha$, active coupling coefficient $\beta$, passive efficiency $\eta$, and internal efficiency $\xi$ (equivalent to the efficiency at the first lens) as a function of disc diameter.
}
\label{fig3}
\end{figure*}

We replace the planar (infinite) metal sheets with finite-size metal discs to confine the light laterally, forming confined OTSs. The efficiency at the first lens, $\xi$, is calculated for a range of discs of different diameters, $D_\text{Disc}$, as summarised in Fig.\ref{fig3}(a). White points denote local maxima, corresponding to the fundamental and higher order modes. As the diameter increases, the structure becomes increasingly multimoded, as expected. For a disc of diameter 2.25$\mu$m, the emission into each decay channel ($P_\text{Absorbed}, P_\text{Top},P_\text{Side},P_\text{Bottom}$) is shown in Fig.\ref{fig3}(b) - the increased emission into the fundamental mode can be seen at 1.3$\mu$m, accompanied by the increase in absorption and suppression of side emission. In order to discuss the far field, we represent the angle of emission $\theta$ relative to the structure (Fig.\ref{fig3}(c)). In our later figures, we integrate the field intensity $|E|^2$ over the azimuthal angle.

The angular (far-field) distribution of the emitted light from the confined Tamm structures with various disc diameters is investigated. The far-field angular distribution is calculated by projecting the electromagnetic field acquired by the top monitor into the far-field by using a near-to-far field transformation. It is essential to engineer the angular distribution when the emitted light is to be coupled to other optics such as an optical fibre or an objective lens. The angular distribution of the incident light must be well within the numerical aperture (NA) of the optical instrument to achieve a high coupling efficiency. In Fig.\ref{fig3}(d)-(g), the far-field distribution of the emission from the confined Tamm structures $|E|^2$ with different disc diameters $D_\text{Disc}$ = 2 - 5$\mu$m are plotted against the emission angle and the wavelength of the light emitted by the dipole source. The numerical aperture of a standard single mode optical fibre SMF-28 (NA = 0.14) is marked by the white lines, and that of a high-NA objective lens (NA = 0.70) is marked by the orange lines. With a small gold disc ($D_\text{Disc} \sim 2 \mu$m), the emitted light from the confined Tamm structure is scattered by the edge of the metal disc into wide angles, as seen in panel (d). The angular distribution of the mode becomes more confined for large gold discs. A parabolic angular distribution (panel (g)) is seen for a very wide (5 $\mu$m) confined Tamm structure, reflecting the transition to a continuum of modes for the increased disc diameter.

In between these two extremes lies a confined Tamm structure with $D_\text{Disc} = 2.25 \mu$m (panel (f)), where the radiation has a well-defined mode and is distributed close to the normal direction. It can be efficiently collected by an objective lens with NA = 0.70 (within orange lines). The emergence of the higher-order modes for large confined Tamm structures can be clearly seen from the far-field distribution. Other than the fundamental mode emitted in the normal direction, additional higher-order modes appear at shorter wavelengths with a wider emission angle.

The absorption in the metal $\alpha$, the calculated $\beta$ factor, the passive emission efficiency $\eta$, and the efficiency at the first lens $\xi$ for the fundamental mode are plotted against the disc diameter of the confined Tamm cavity in Fig.\ref{fig3}(h). We can see that $\beta$ (red) increases with the disc diameter, as less light is scattered to the side for a wide confined Tamm structure. However, $\eta$ (green) decreases with increasing disc diameter, indicating that more light is emitted into the substrate, compared to the top surface. Moreover, the absorption in the gold disc $\alpha$ (grey) also increases. These three factors compete and result in no significant change in the extraction efficiency $\xi$ (blue) with respect to the disc diameters for larger disc size. $D_\text{Disc} = 2.25 \mu$m chosen as the preferred disc diameter accounting for both the extraction efficiency (Fig.\ref{fig3}(h)) and the far-field distribution (Fig.\ref{fig3}(d)-(g)). For this disc diameter of 2.25 $\mu$m, the first-lens efficiency of the confined Tamm structure $\xi$ is 18\%, with a corresponding $\alpha$ of $\sim40\%$.

\section*{Nanoring Tamm Design Principles}

\begin{figure*}
\includegraphics[width=\linewidth]{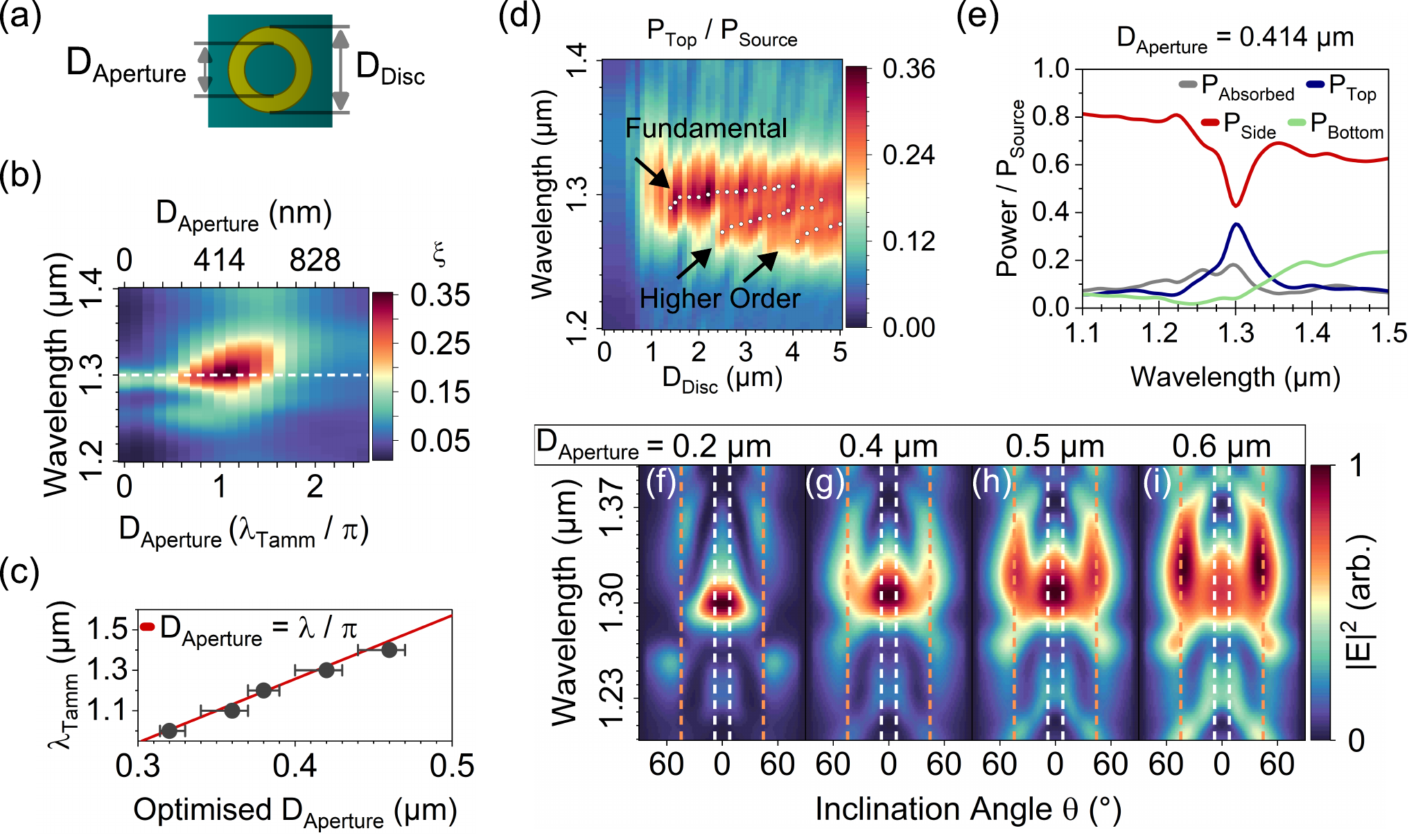}
\caption
{Summary of FDTD calculations for Tamm nanorings. 
(a) A birds-eye view illustration of a Tamm nanoring structure with external diameter $D_\text{Disc}$ and internal diameter $D_\text{Aperture}$. (b) The power transmission through the top surface for a Tamm nanoring with a fixed $D_\text{Disc} = 2.25$ µm and a variable aperture diameter. Compared to a confined Tamm structure with a disc of the same outer diameter, $P_\text{Top}/P_\text{Source}$ increases from 16\% (no aperture) to 36\% for an optimised inner diameter $D_\text{Aperture}$. (c) For Tamm nanoring devices designed for a range of wavelengths $\lambda_\text{Tamm} = 1-1.5 \mu m$, the aperture diameter which maximises the outcoupling efficiency $\xi$ have been plotted in black. The red line illustrates a plot of $\lambda/\pi$ (see text for discussion). (d) Calculated power transmission through the top of the nanorings as a function of the outer diameter with an optimised inner diameter $D_\text{Aperture} = \lambda/\pi$. Local maxima are indicated with white dots to serve as a visual guide for the fundamental and higher order modes. (e) The absorption and power transmission of a $D_\text{Disc} = 2.25 \mu m$ and $D_\text{Aperture} = 0.414 \mu m$ nanoring show that the addition of an aperture causes significant reduction of absorption and consequently a near doubling of the $P_\text{Top}/P_\text{Source}$ efficiency as compared to the analogous disc confined Tamm in Fig.\ref{fig3}(b). (f)-(i) The far-field projections of the emitted E-field across a range of inner diameters $D_\text{Aperture}$.
}
\label{fig4}
\end{figure*}

To reduce this substantial absorption, we introduce an aperture in the gold disc to form a nanoring structure, as schematically illustrated in Fig.\ref{fig4}(a). Using our optimised design thus far ($D_\text{Disc} = 2.25 \mu$m, $d_\text{Metal}$ = 20 nm, $d_\text{Spacer}$ = 64 nm) as a starting point, we calculate the efficiency for a range of structures with varying inner radius $D_\text{Aperture}$; this is shown as a wavelength resolved plot in Fig.\ref{fig4}(b) where the diameter has been shown normalised to the Tamm wavelength $\lambda_\text{Tamm}$.

 A clear enhancement in the extraction efficiency is observed when the aperture diameter is smaller than 800 nm. At $D_\text{Aperture}$ = 414 nm, the efficiency $\xi$ is maximised. This appears close to $\lambda_{Tamm}$ /$\pi$. Remarkably, along with the aperture diameter, the efficiency improves by a factor of almost two from 18\% to 35\%. Simultaneously, $\lambda_\text{Tamm}$ red-shifts from 1.296 $\mu$m to 1.325 $\mu$m when the $D_\text{Aperture}$ increases to 0.8 $\mu$m.

To validate the conjecture that this optimal aperture diameter is located at $D_\text{Aperture}$ = $\lambda_\text{Tamm}$ /$\pi$, we simulate the extraction efficiency of nanoring Tamm structures designed for a range of $\lambda_\text{Tamm}$ = 1-1.5$\mu$m.
 To achieve different centre wavelengths, the thickness of the DBR layers and the spacer layer are appropriately varied, while the metal layer thickness is fixed to be 20 nm. The FDTD simulations are run for each nanoring Tamm structure by varying the aperture diameters from 0 to 1 $\mu$m, with the outer $D_\text{Disc}$ fixed at 2.25$\mu$m. Figure \ref{fig4}(c) shows the optimised aperture diameter for each $\lambda_\text{Tamm}$. The optimised $D_\text{Aperture}$ values agree well with $\lambda_\text{Tamm}$ = $\pi$ $D_\text{Aperture}$, which is shown as a red line. We note that $\xi$ is maximised for these different structures for aperture circumferences equal to the Tamm wavelength, which suggests hybridised annular modes from the ring.

Finally, we investigate the Tamm nanoring structure by varying the outer disc diameter $D_\text{Disc}$, while keeping $D_\text{Aperture}$ = 414 nm. The efficiency is illustrated in Fig.\ref{fig4}(d).
The modes red-shift and the structure becomes multimoded with increasing disc diameter, equivalent to the solid metal disc confined Tamm modes.

This shows the modes are only slightly perturbed by the introduction of the aperture, and that the enhancement of the emission efficiency may be caused by the hybridisation of the confined Tamm mode and the annular local mode in the aperture. Applying the 414nm ($\approx\lambda_\text{Tamm}/\pi$) aperture to the structure described above, we present the power transmissions $P_\text{Top},P_\text{Side}, $ and $P_\text{Bottom}$, and the absorbed power $P_\text{Absorbed}$ for the nanoring Tamm in Fig.\ref{fig4}(e). It can be seen that at the central $\lambda_\text{Tamm} = 1.3\mu m$, the efficiency at the first lens is nearly double ($\xi=35\%$) compared to that of the regular solid disc, owing to the significantly lower absorption $\alpha = 18\%$.

We check for any aperture-induced changes on the angular distribution of the field reaching the first lens of nanoring structures with a fixed outer $D_\text{Disc} = 2.25\mu m$ and a varying inner $D_\text{Aperture}$, as shown in Fig.\ref{fig4}(f)-(i).
While the aperture does induce a minor impact of the far field distribution, the majority of the emission remains within the 0.7 NA as for the solid disc considered previously.

\section*{Conclusion}

Tamm structures are an intriguing form of readily manufactureable photonic cavity which are potentially suitable for both quantum and classical sources of light. However, absorption losses in the metal challenge the viability of real-world Tamm devices. 
In this paper, we have proposed a nanoring Tamm structure as a mechanism to  increase the efficiency by reducing this absorption while maintaining a plausible far-field distribution for mode-coupling such devices.

We have shown that, for a design optimised for the telecommunications O band, replacing a traditional confined Tamm structure with a nanoring can nearly double the output power transmission, provided the aperture of the nanoring is appropriately optimised. For our design, a confined Tamm structure with outer $D_\text{Disc}$ = 2.25$\mu$m results in $\xi=18\%$. Introducing an aperture of $D_\text{Aperture} = \lambda_\text{Tamm}/\pi$ (414 nm) increases $\xi$ to $35\%$. Such an efficiency is sufficiently high for a SPS for quantum communications, and combines brightness with a highly manufactureable structure. We believe this will motivate fabrication and further research of such devices.

\section*{Author Contributions}
H.H. established the idea, performed the simulations with assistance from D.D. and J.R.P., performed data analysis and drafted the paper. D.D performed supplementary simulation. E.H and R.O supervised the project. All authors were involved in the discussion and contributed to the final version of the paper. 

\begin{acknowledgement}

HH thanks the Quantum Engineering Centre for Doctoral Training, and all authors acknowledge funding from EPSRC under grants META TAMM (EP/X029360/1), SPIN SPACE (EP/M024156/1), 1D QED (EP/N003381/1) in support of this research. We also acknowledge the Advanced Computing Research Centre in University of Bristol for offering the computational resources to perform the simulations.

\end{acknowledgement}

\bibliography{Tamm reference}

\end{document}